\documentclass[aps,pra,twocolumn,superscriptaddress]{revtex4-2}
\usepackage{graphicx}
\usepackage{url}
\usepackage{amsmath,amsthm,amssymb,physics,enumerate,tabularx,makecell,mathrsfs,float,bbm,stackrel,mathtools,multirow,bm}
\usepackage{xcolor}
\usepackage{natbib}
\usepackage{physics}
\usepackage{chemformula}

\usepackage[colorlinks=true,%
bookmarks=false,%
linkcolor=blue,%
urlcolor=blue,%
citecolor=blue,%
breaklinks]{hyperref}
\usepackage{tabularx}

\begin{document}
    \title{Estimating molecular thermal averages with the quantum equation of motion and informationally complete measurements}

\author{Daniele Morrone}
\email{daniele.morrone@unimi.it}
\affiliation{Quantum Technology Lab, Dipartimento di Fisica Aldo Pontremoli,
Università degli Studi di Milano, I-20133 Milano, Italy}
\affiliation{Algorithmiq Ltd, Kanavakatu 3C 00160 Helsinki, Finland}

\author{N. Walter Talarico}
\affiliation{Algorithmiq Ltd, Kanavakatu 3C 00160 Helsinki, Finland}
\affiliation{HelTeq group, QTF Centre of Excellence, Department of Physics, University of Helsinki, P.O. Box 43, FI-00014 Helsinki, Finland.}

\author{Marco Cattaneo}
\affiliation{Algorithmiq Ltd, Kanavakatu 3C 00160 Helsinki, Finland}
\affiliation{HelTeq group, QTF Centre of Excellence, Department of Physics, University of Helsinki, P.O. Box 43, FI-00014 Helsinki, Finland.}
\affiliation{Pico group, QTF Centre of Excellence, Department of Applied Physics,
Aalto University, P.O. Box 15100, FI-00076 Aalto, Finland}

\author{Matteo A. C. Rossi}
\affiliation{Algorithmiq Ltd, Kanavakatu 3C 00160 Helsinki, Finland}
\affiliation{HelTeq group, QTF Centre of Excellence, Department of Physics, University of Helsinki, P.O. Box 43, FI-00014 Helsinki, Finland.}
\affiliation{Pico group, QTF Centre of Excellence, Department of Applied Physics,
Aalto University, P.O. Box 15100, FI-00076 Aalto, Finland}

\date{\today}
	
\begin{abstract}
By leveraging the Variational Quantum Eigensolver (VQE), the ``quantum equation of motion" (qEOM) method established itself as a promising tool for quantum chemistry on near term quantum computers, and has been used extensively to estimate molecular excited states. Here, we explore a novel application of this method, employing it to compute thermal averages of quantum systems, specifically molecules like ethylene and butadiene.
A drawback of qEOM is that it requires measuring the expectation values of a large number of observables on the ground state of the system, and the number of necessary measurements can become a bottleneck of the method. In this work we focus on measurements through informationally complete positive operator-valued measures (IC-POVMs) to achieve a reduction in the measurements overheads. We show with numerical simulations that the qEOM combined with IC-POVM measurements ensures a satisfactory accuracy in the reconstruction of the thermal state with a reasonable number of shots.
\end{abstract}

\maketitle
\section{Introduction}
\label{s:Introduction}
Computing properties of quantum systems at finite temperature poses a significant challenge due to its inherent resource-intensive nature. The computational cost of such calculations generally scales exponentially with the system size, unless symmetries or other characteristics of specific systems can be leveraged. Efficient classical computation of thermal averages is therefore strongly subjected to the type of system one is interested in. 
The construction and estimation of properties of thermal states is a task required in different fields, ranging from quantum chemistry \cite{QuantumChemistry2020} and many-body physics \cite{QuantumMany2023} to high energy physics \cite{Wiese2013}. 
Classical methods typically used to evaluate such quantities include algorithms from the quantum Monte Carlo family \cite{boothApproachingChemicalAccuracy2010,tubmanDeterministicAlternativeFull2016}, Density Matrix Renormalization group (DMRG) \cite{knechtNewApproachesInitio2016}, quantum belief propagation \cite{hastingsQuantumBeliefPropagation2007} and minimally entangled typical thermal states (METTS)\cite{stoudenmireMinimallyEntangledTypical2010,whiteMinimallyEntangledTypical2009}.

In recent years, new types of quantum algorithms including adaptations of classical ones have been developed to leverage the capacity of quantum computers \cite{nielsen_chuang_2010}, both for the near-term and fault-tolerant eras \cite{lau2022nisq}. Algorithms developed for fault-tolerant quantum computers typically require the use of Quantum Phase Estimation (QPE) \cite{temme_quantum_2011,yung_quantumquantum_2012,poulin_sampling_2009} as a sub-routine, while those made for near-term devices are typically based on variational routines such as the Variational Quantum Eigensolver (VQE) \cite{selisko_extending_2022,verdon2019quantum} or Quantum Approximate Optimization Algorithms (QAOA) \cite{wu_variational_2019,sagastizabal_variational_2021}. 

In this work, we focus on molecular systems such as linear polyenes \cite{knechtPhotophysicsCarotenoidsMultireference2013,christensenEnergiesLowLyingExcited2008,tavan1Ag1BuEnergy2008,christensenLinearPolyenesModels2004,krawczykVibronicStructureCoupling2013,CarotenoidExcited2005} as testbeds over which reconstruct an approximate low temperature thermal state by means of variational quantum algorithms.
Among the methods based on the use of the VQE, the Quantum Equation Of Motion (qEOM)  \cite{ollitrault_quantum_2020,rizzo_one-particle_2022} is a hybrid method that allows for the evaluation of excited states of a system. More generally, the qEOM belongs to a class of methods referred to as Quantum Subspace Methods (QSM)\cite{motta2023subspace,urbanek2020,takeshita2020,Stair2020,Cortes2022}, which are becoming increasingly popular for both near-term and fault-tolerant quantum computing for the exploration of the low- and the high-lying excited state, energies, as well as molecular properties \cite{gandon2024nonadiabatic,reinholdt2024subspace,jensen2024quantum}.

\begin{figure*}
		\includegraphics[width=0.95\textwidth]{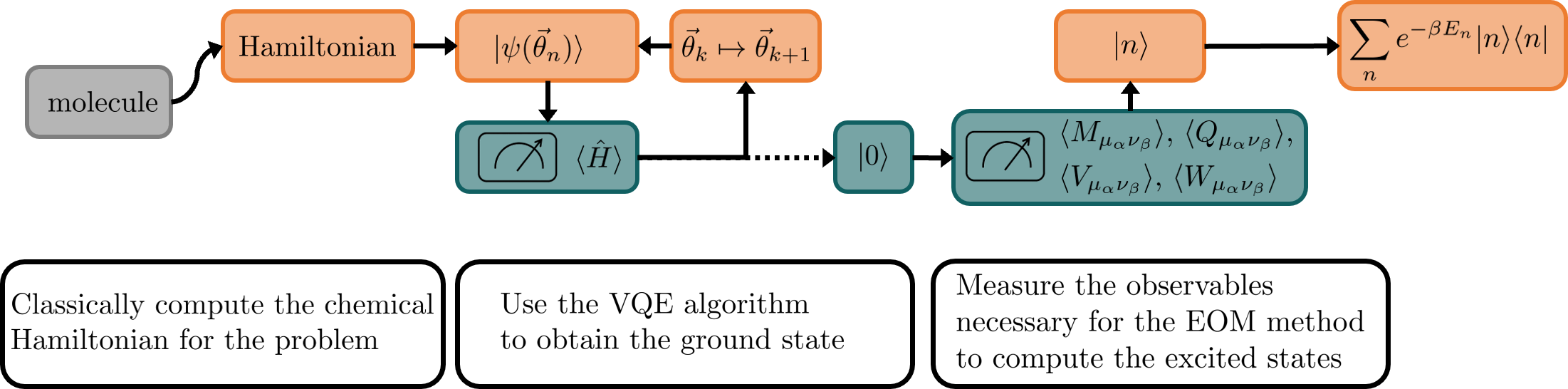} 
	\caption{Representation of the qEOM method. This hybrid method requires calculation performed both on a classical calculator (orange boxes) and on a quantum processor (green boxes). The scheme also depict that information is frequently passed between the two devices. The various steps of the method are described in the white boxes.} 
	\label{f:schematics}
 \end{figure*}

The "Equation Of Motion" is a classical chemistry algorithm \cite{Rowe} used to infer a finite number of excited states for a molecular system. To do so, it constructs a subspace of states with a limited number of excitations, which can be done by applying excitation operators onto the ground state (GS), from which the eigenstates of the Hamiltonian included in the subspace can be found by minimizing the energy. In this way, a Generalized Eigenvalue Problem (GEP) is formed, which can be solved to recover the eigenstates. In their work \cite{ollitrault_quantum_2020} \textit{Ollitrault et al.} introduced the hybrid version of the methods, where the GS is prepared on a quantum computer and the terms of the GEP are measured directly from it. Doing so improves the efficiency in preparing the GS \cite{peruzzoVariationalEigenvalueSolver2014,QuantumChemistry2020}.  

A crucial aspect of the optimal usage of near-term devices is the choice of measurement scheme. Indeed, while in the fault-tolerant era an efficient measurement scheme based on the QPE is known \cite{KnillOptimalQuantumMeasurements2007}, in the near-term era it is not trivial to find an optimal measurement strategy, as different methods may produce different results when it comes to number of required measurements, circuits depths and accuracy. Promising methods involves many ideas to optimize the measurement process, such as grouping of commuting observables for simultaneous measuring \cite{kandalaHardwareefficientVariationalQuantum2017, yen2020measuring,hugginsEfficientNoiseResilient2021,Bonet2020NearlyOptimal,cotlerQuantumOverlappingTomography2020}, classical shadows \cite{huangPredictingManyProperties2020,hadfield2020measurements,EfficientEstimation2021}, and machine learning assisted tomography \cite{torlaiNeuralnetworkQuantumState2018}.
In combination with the qEOM methods, we will perform measurements using informationally complete positive operator-valued measures (IC-POVMs), which have been recently demonstrated to be a general framework to efficiently estimate many-body operators on near-term devices \cite{GarcíaPérezGuillermo2021LtMA}. The qEOM methods requires to measure a number of observable that grows rapidly with the system size. In such cases, IC-POVMs allow for the efficient estimation of different observables using a scalable number of shots. Furthermore, IC-POVMs may be adapted variationally to further lower the variance of the estimation of the observables of interest, i.e., to reduce the number of measurement shots \cite{GarcíaPérezGuillermo2021LtMA}.

This work is organized as follow: in Sec. \ref{s:qeom} we review the (q)EOM methods. In Sec. \ref{s:povm} we review the  IC-POVM measurement scheme. In Sec  \ref{s:res} we present our result and in Sec. \ref{s:conc} we give our outlook and conclusion.


\section{Quantum Equation of motion}
\label{s:qeom}

In this section, we provide a brief overview of both classical and quantum versions of the Equation of Motion (EOM) method for computing electronic excitation energies \cite{ollitrault_quantum_2020, rizzo_one-particle_2022}. The primary goal of the EOM approach is to derive excited states of a molecule from its GS, represented as $|0\rangle$. This is achieved by constructing and solving an eigenvalue problem within a reduced subspace, limiting the size of the problem to be manageable by conventional classical algorithms. The construction of this reduced subspace involves spanning a set of states derived from applying various excitation operators, including both single and multiple excitations, onto the GS.

In the EOM method we assume an approximate understanding of the GS of the system, denoted as $\ket{0}$. The difference between classical and quantum EOM methods lies in how we determine $\ket{0}$. Classically, the GS can be inferred using various methods developed for such purpose. However, attaining the desired accuracy swiftly becomes computationally prohibitive as the size of the molecule increases. This is primarily due to the limitations in scaling that classical computational methods encounter. To enhance efficiency and achieve superior scaling, it becomes imperative to transcend classical computation and harness quantum effects \cite{abrams1997,abrams1999,guzik2005,peruzzoVariationalEigenvalueSolver2014}.

A quantum computer may be exploited to obtain a more efficient and faithful GS of the molecule. To achieve this, the fermionic problem (i.e., the second-quantized Hamiltonian of the molecule) must first be transformed into a qubit problem using established mappings \cite{BRAVYI2002210, Jiang_2020, bravyi2017tapering, setia2018}. Then, the Hamiltonian can be represented using gates on a quantum computer, and the GS can be prepared, for instance, by employing the Variational Quantum Eigensolver (VQE) algorithm \cite{mccleanTheoryVariationalHybrid2016, peruzzoVariationalEigenvalueSolver2014}.

Let us introduce the generic excitation operator as $\hat{O}_n^\dagger=|n\rangle \langle 0 |$, where $\ket{n}$ is the $n$th excited state of the molecule. Analogously, $O_n= \ket{0}\bra{n}$. We can readily observe that the expectation value for the excitation energies, denoted as $E_{0n}=E_n-E_0$, can be expressed as \cite{ollitrault_quantum_2020}:
\begin{align}
	E_{0n}=\frac{\langle 0| \left[ \hat{O}_n,\left[ \hat{H},\hat{O}_n^\dagger \right]  \right]|0\rangle}{\langle 0|\left[ \hat{O}_n, \hat{O}_n^\dagger \right] |0\rangle}. \label{e:excitation_op}
\end{align}
The key idea of the EOM method lies in the representation of the generic excitation operator $O_n$. We choose to write it as a linear combination of a finite number of single and double excitations denoted by $\hat{E}_{\mu_\alpha}^{(\alpha)}$ 
\begin{align}
	\hat{O}_n^\dagger = \sum_\alpha \sum_{\mu_\alpha}\left[
	X_{\mu_\alpha}^{(\alpha)} (n) \hat{E}_{\mu_\alpha}^{(\alpha)} - Y_{\mu_\alpha}^{(\alpha)} (n) (\hat{E}_{\mu_\alpha}^{(\alpha)})^\dagger \right]\label{e:expansion_op}
\end{align}
where $\alpha$ refers to the number of excitations considered ($\alpha=1$ for single excitations and $\alpha=2$ for double excitations), $\mu_\alpha$ is a collective index for all one-electron orbitals involved in the excitation, and $X_{\mu_\alpha}^{(\alpha)} (n)$, $Y_{\mu_\alpha}^{(\alpha)} (n)$ are the coefficients of the linear combination. For instance, a single excitation is represented by $\hat{E}_{\mu_1}^{(1)}=a_m^\dagger a_i$ for some orbitals $m$ and $i$. The span of all the excitations $\hat{E}_{\mu_\alpha}^{(\alpha)}$ applied onto the GS forms the subspace over which we solve the EOM problem. 

We can finally insert the expansion in Eq.~\eqref{e:excitation_op} to obtain a parametric equation for the coefficients of Eq.~\eqref{e:expansion_op}. To find the approximate eigenvalue of the Hamiltonian in the generated subspace, and identify the coefficients of the linear expansion, we need to look for minima of the energy in the coefficients space. We use the variational principle for local minima and evaluate $\delta(E_{0n})=0$. By doing so, we obtain a Generalized Eigenvalue Problem (GEP) whose solutions lead to the coefficients we are interested in. The GEP reads:
\begin{align}
	\begin{pmatrix}
		M & Q \\
		Q* & M* 
	\end{pmatrix}\begin{pmatrix}
		X_n \\
		Y_n 
	\end{pmatrix}= E_{0n} \begin{pmatrix}
		V & W \\
		-W* & -V* 
	\end{pmatrix}
	\begin{pmatrix}
		X_n \\
		Y_n 
	\end{pmatrix},\label{e:GEP}
\end{align}
where the matrix elements can be expressed as expectation values over the GS $|0\rangle$:
\begin{align}
	M_{\mu_\alpha\nu_\beta}=&\langle 0 | \left[(\hat{E}_{\mu_\alpha}^{(\alpha)})^\dagger,\hat{H},\hat{E}_{\nu_\beta}^{(\beta)}\right] |0 \rangle, \\
	Q_{\mu_\alpha\nu_\beta}=&-\langle 0 | \left[(\hat{E}_{\mu_\alpha}^{(\alpha)})^\dagger,\hat{H},(\hat{E}_{\nu_\beta}^{(\beta)})^\dagger \right] |0 \rangle, \\
	V_{\mu_\alpha\nu_\beta}=&\langle 0 | \left[(\hat{E}_{\mu_\alpha}^{(\alpha)})^\dagger,\hat{E}_{\nu_\beta}^{(\beta)}\right] |0 \rangle, \\
	W_{\mu_\alpha\nu_\beta}=&-\langle 0 | \left[(\hat{E}_{\mu_\alpha}^{(\alpha)})^\dagger,(\hat{E}_{\nu_\beta}^{(\beta)})^\dagger \right] |0 \rangle,
\end{align}
with the double commutator defined as $[\hat{A},\hat{B},\hat{C}]= \{[[\hat{A},\hat{B}],\hat{C}]+[\hat{A},[\hat{B},\hat{C}]]\}/2$.
 
In the classical EOM method, the elements of the GEP are numerically computed from the approximate GS. Similarly, in the qEOM variant, the elements are obtained by performing measurements on the GS prepared on a quantum computer. Note that the number of observables that need to be estimated grows rapidly with the system size. Then, measuring with IC-POVMs a ground state prepared on a quantum computer is much more efficient than computing classically all the observables of interest, which quickly becomes unfeasible for large molecules. Once the GEP has been reconstructed, for both methods, it can be solved through classical algorithms to find a set of excited states of the molecule.

\section{Informationally-Complete POVMs}
\label{s:povm}

In the qEOM method, once the state of the quantum computer has been evolved to the GS of the system of interest, it is necessary to recover the information needed to reconstruct the EOM matrices. Full Quantum State Tomography (QST) is the method used to characterize unknown quantum states, but it is particularly resource intensive. In most cases, the full knowledge of the state is not necessary, so further reduction in the required resource  can be achieved by using techniques that allow for partial state tomography, such as classical shadows \cite{Aaronson2018,huangPredictingManyProperties2020} and, more generally, informationally complete POVMs. 
In our work we use local single-qubit IC-POVMs to reconstruct the EOM matrices. In this section we review how such measurements are performed according to this method, as described in \cite{GarcíaPérezGuillermo2021LtMA,Acharya2021}.

A POVM is said informationally complete if its effects are linearly independent and span the space of linear operators in the Hilbert space of the system. The number of effects of the POVM needs to be at least $d^2$ where $d$ is the dimension of the Hilbert space. For a qubit this means 4 effects. For a system of $N$ qubits, $4^N$.

An operator $ \hat{O}$ can be decomposed in the basis of the effects of a given POVM $\{ \Pi_m\}$:
\begin{align}
	\hat{O}=\sum_m \omega_m \Pi_m,
\end{align}
 which makes its expectation value over the state $\rho$
 \begin{align}
 	\langle \hat{O} \rangle = \Tr\left[ \rho \hat{O} \right] = \sum_m \omega_m p_m ,\label{e:expectation_value}
 \end{align}
where $p_m = \Tr[\rho\Pi_m]$ is the probability of the outcome $m$ to be measured.

The first step of the method is to construct a stochastic estimator for the expectation value by performing a Monte Carlo sampling. Specifically, by sequentially measuring the system of a quantum computer $S$ times it is possible to obtain a sequence of outcomes $m_s$ from the distribution $\{ p_m \}$. The weight $\omega_m$ can be classically computed in polynomial time. This sequence of samples can then be used to construct the estimator
 \begin{align}
	\bar{O} = \Tr\left[ \rho \hat{O} \right] = \frac{1}{S} \sum_{s=1}^{S} \omega_{m_s}.
\end{align}
The statistical error for the procedure is \cite{GarcíaPérezGuillermo2021LtMA}
\begin{align}
\sqrt{\frac{\langle \omega_m^2 \rangle _{p_m} - \langle \omega _m \rangle_{p_m}^2 }
	{S}}
\end{align}

Ref.~\cite{GarcíaPérezGuillermo2021LtMA} proposes to update the choice of the IC-POVM used in the measurement procedure so that the above-defined estimator has minimal statistical error. In our work we do not perform this optimization process, while we employ a symmetric IC-POVM (SIC-POVM), which is a reasonable choice in absence of prior knowledge about the state of the system \cite{GarcíaPérezGuillermo2021LtMA}. Symmetric POVM means that the single qubit effects, when rescaled as $\Tilde{\Pi}_i=2\Pi_i$, fulfill the relation $\Tr\left[ \Tilde{\Pi}_i \Tilde{\Pi}_j  \right]=(2\delta_{ij}+1)/3, \; \forall i,j$. Specifically, we consider the canonical SIC POVM $\{ \tilde\Pi_i = |\tilde\pi_i\rangle\langle\tilde\pi_i|\}_{i=0}^3$  given by the projectors onto $\ket{\tilde\pi_0} = \ket{0}$, $\ket{\tilde\pi_k} = \ket{0} + \sqrt{2}e^{i2\pi(k-1)/3}\ket{1}$ for $k\in \{1, 2, 3\}$.

We leave the assessment of the performance of adaptive measurements to future work.

\section{Results}
\label{s:res}
\begin{figure*}
	\begin{tabularx}{2\columnwidth}{XX}
		(a)
		
		\includegraphics[width=0.45\textwidth]{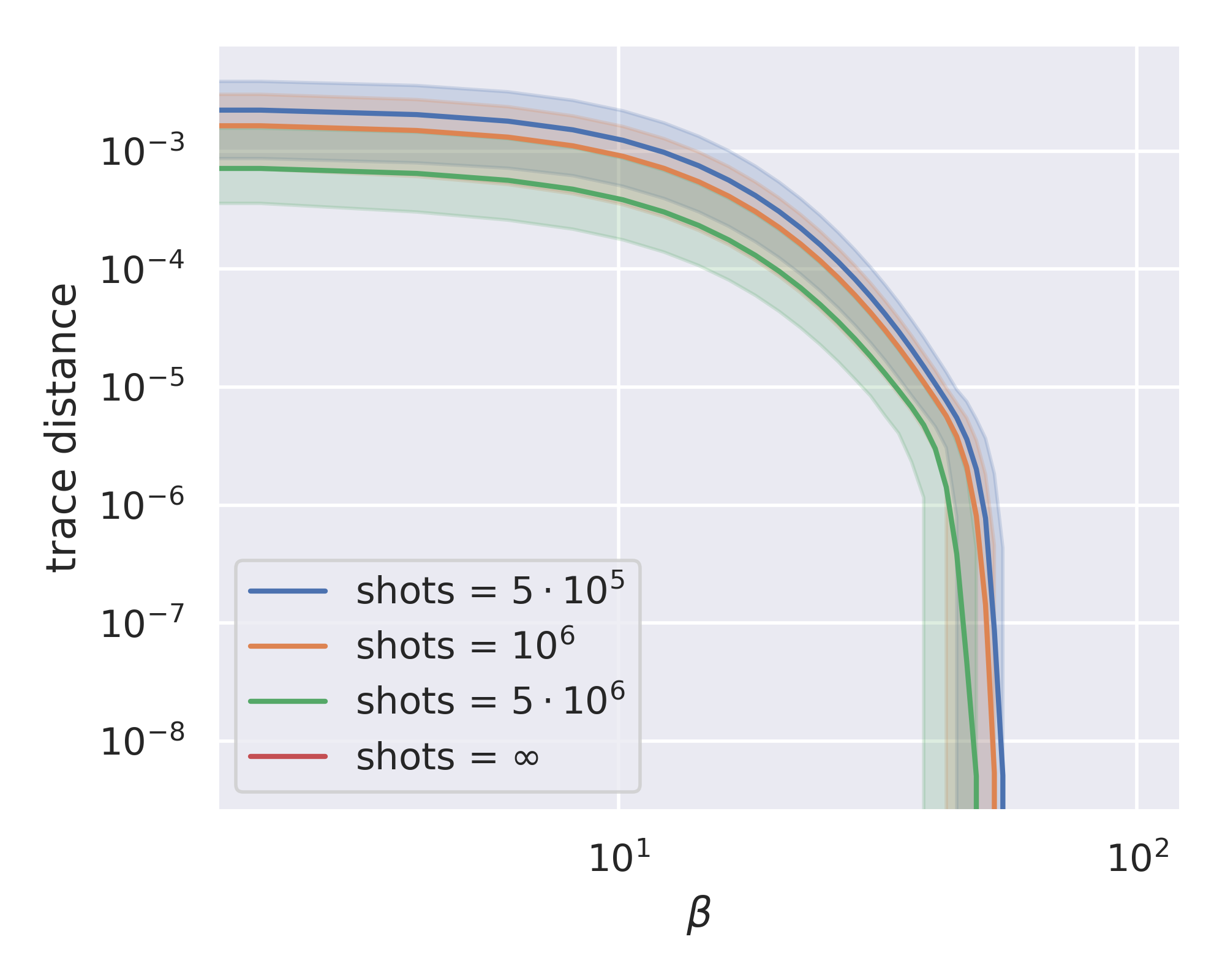}  &
		(b)
		
		\includegraphics[width=0.45\textwidth]{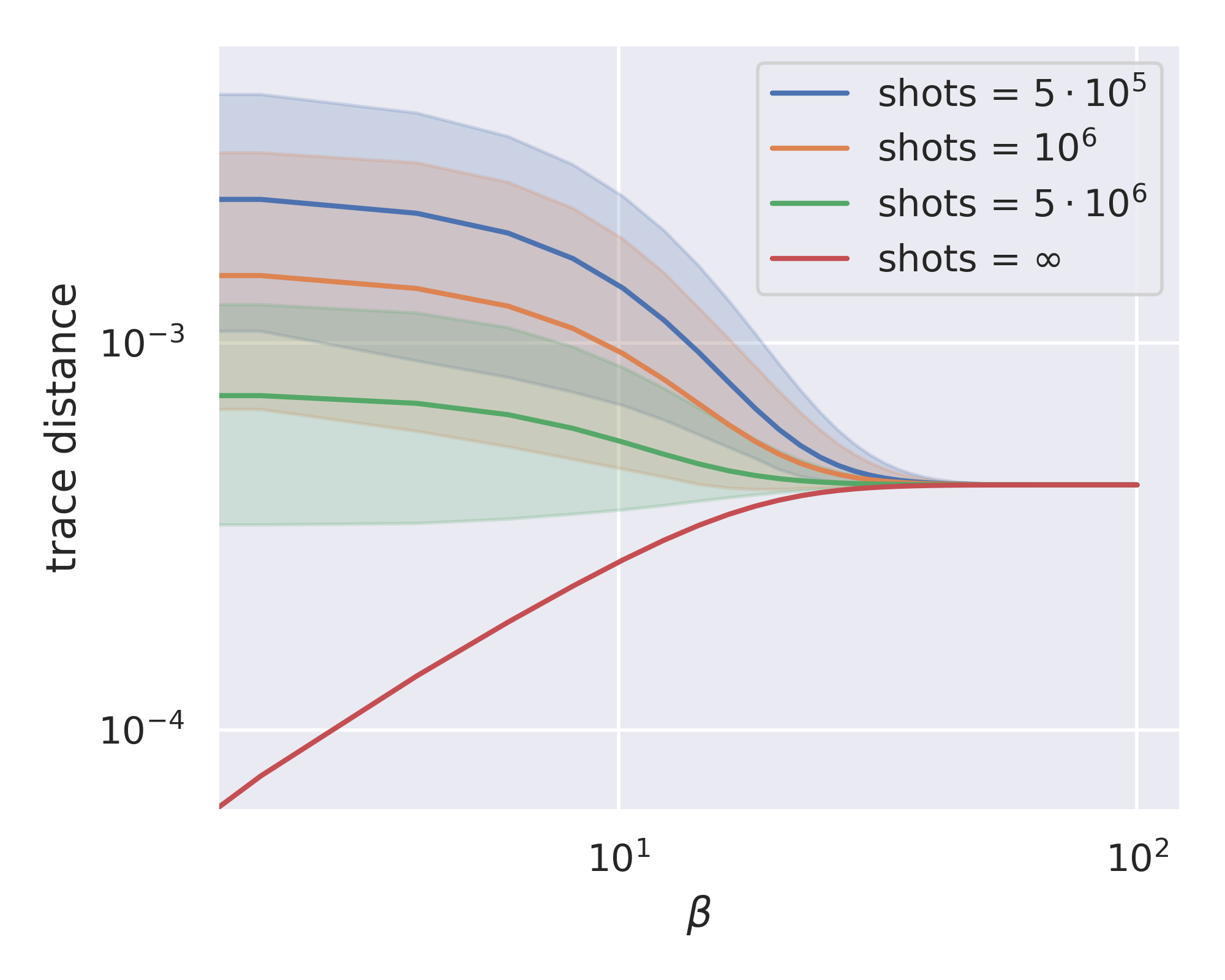}  \\
		(c)
		
		\includegraphics[width=0.45\textwidth]{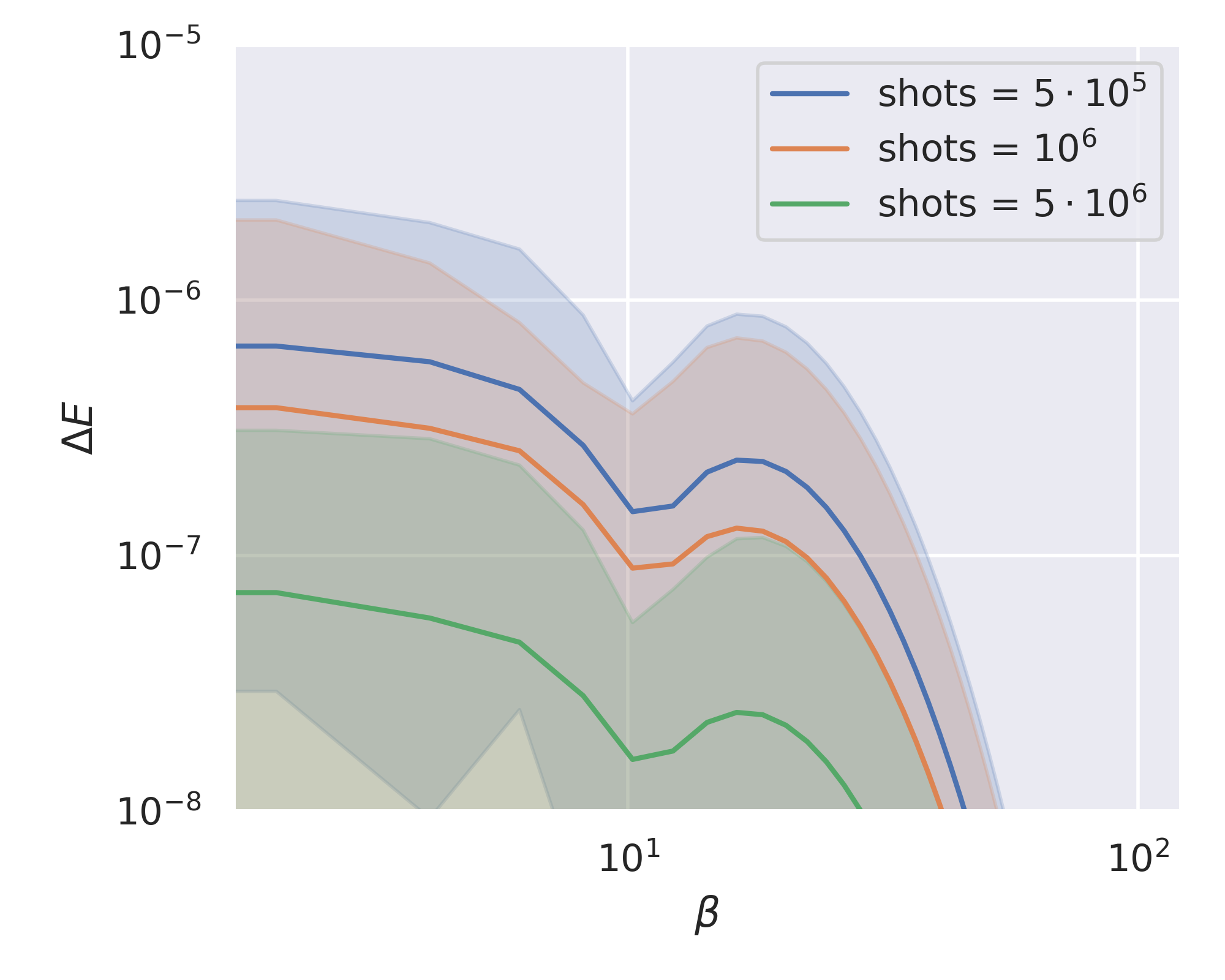}  &
		(d)
		
		\includegraphics[width=0.45\textwidth]{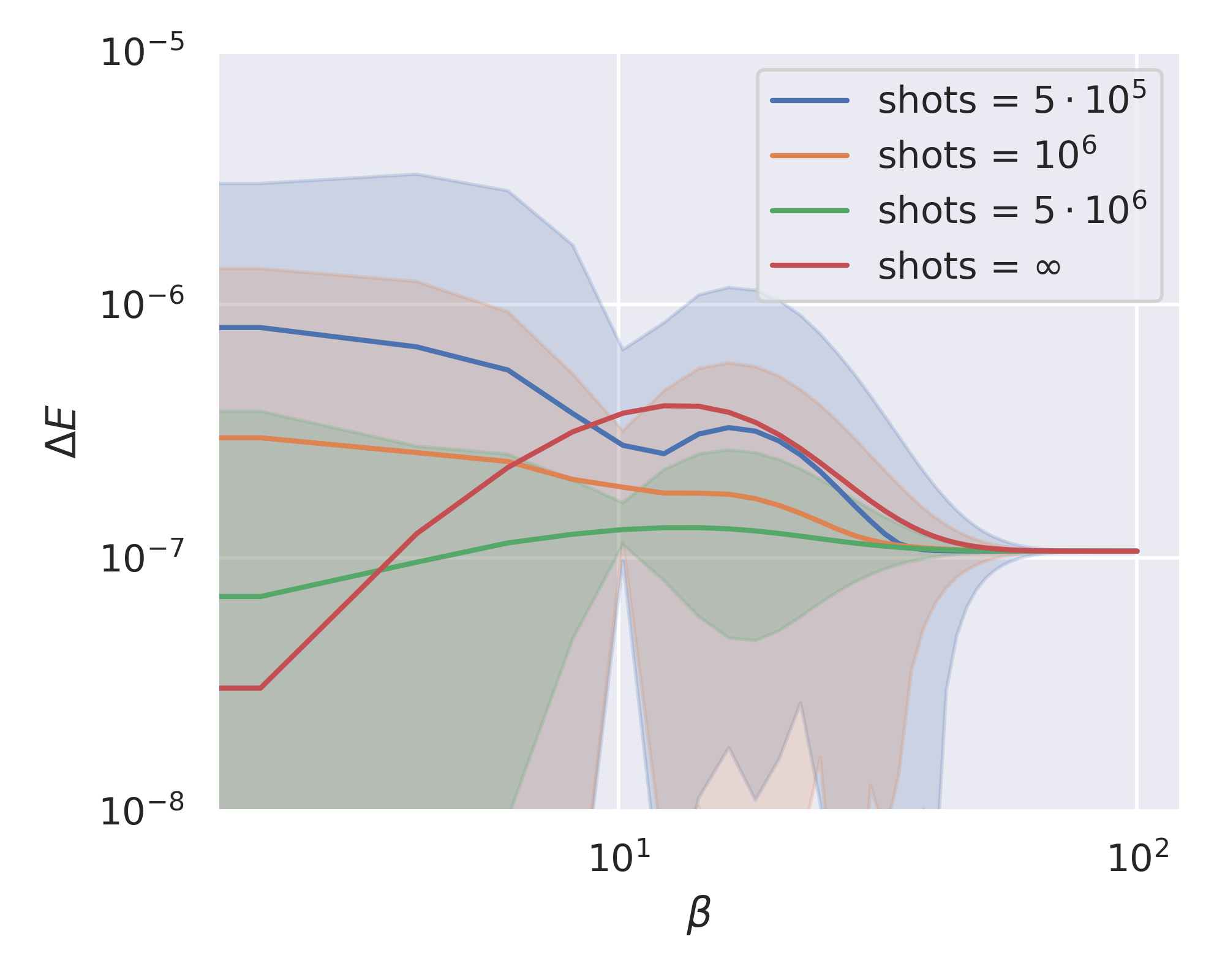}  \\
	\end{tabularx}
	\caption{Ethylene: Top row, trace distance between ideal thermal state and reconstructed thermal state through the qEOM method for different number of shots and starting from the exact ground state (panel a) and from a ground state found through a VQE (panel b). Bottom row, error for the average energy for the same states as above. All simulations run with finite sampling are repeated 100 times, and the spread of results is shown in the figures as an interval with a percentile width of 99.7.} 
	\label{f:ethylene}
 \end{figure*}

In this section, we present results from simulating the qEOM algorithm on a classical computer. We evaluate the algorithm's performance on linear polyenes with increasing chain lengths, specifically ethylene and butadiene (\ch{C_{2n}H_{2n+2}} with $n=1,2$). A correlation-consistent (cc-pVDZ) basis set \cite{dunning1989} is used for both systems. For each molecule, we obtain molecular orbitals and one- and two-electron integrals using the Hartree-Fock method with the quantum chemistry package PySCF \cite{Sun2018,sun2020}. Furthermore, we use the same library for subsequent multiconfigurational self-consistent field (MCSCF) calculations \cite{schmidt1998}. This allows us to select a subset of active space orbitals to further reduce the size of the systems considered, specifically we choose CAS(2,2) for ethylene and CAS(4,4) for butadiene, with a complete active space CAS($n_e$, $n_o$) of $n_e$ electrons in $n_o$ orbitals.  We then utilize Qiskit \cite{Qiskit} to transform the second-quantized fermionic molecular Hamiltonian to a bosonic problem, via the Jordan-Wigner mapping, and perform a VQE routine to find the approximate GS needed for the qEOM algorithm. On top of that, we use the IC-POVM as implemented in the software Aurora \cite{Aurora} to simulate the measurement outcomes of the quantum experiment and to estimate the matrix elements required for the construction of the GEP. Once the eigenstates are obtained, further manipulations are performed with the QuTiP library \cite{qutip1,qutip2}.

For comparison purposes, we also diagonalize the Hamiltonian, filtering the eigenstates with a defined particle number \cite{ollitrault_quantum_2020}, and obtain the exact thermal states for the ideal molecular systems, written as:
\begin{align}
\rho_{th}=\sum_n e^{-\beta E_n} |n \rangle \langle n |.
\end{align}
For the qEOM algorithm, simulations are performed with both the ideal GS obtained from the diagonalization and an approximate GS obtained through the VQE. The VQE has been run with the UCCSD ansatz \cite{sokolov2020,barkoutsos2018} and assuming we are using a noiseless quantum computer. The parameters optimized in the VQE are 3 for the ethylene and 26 for the butadiene. Then, we use IC-POVMs to estimate the observables needed to reconstruct the EOM matrices for both molecules. The number of observables we need to estimate is $100$ for the ethylene and $10816$ for the butadiene. Simulations are repeated multiple times to account for statistical errors due to shot noise. Additionally, for the sake of comparison, we also show the unrealistic case of simulations of qEOM with infinite sampling. We demonstrate the agreement between the ideal thermal state and the thermal state reconstructed with the qEOM method using the trace distance between them as a figure of merit \cite{nielsen_chuang_2010}:
\begin{align*}
D(\rho,\sigma)=\frac{1}{2} \Tr \left[ \sqrt{(\rho-\sigma)^\dagger(\rho-\sigma)} \right].
\end{align*}
Moreover, we also calculate the absolute value of the energy difference between ideal and reconstructed thermal states, $\Delta E$.

For ethylene, as shown in Fig. \ref{f:ethylene}, the method achieves a good approximation regardless of the temperature. The trace distance between the exact thermal state and the thermal state found through the qEOM method, when starting from the exact GS and using infinite sampling, is within numerical error. Noise from limited sampling is more dominant at higher temperatures, where excited states have a heavier weight. 

While not reported, statistics comparing trace distance spread for the thermal state in the case where the GS is exact and for the VQE-approximated GS do not display any noticeable difference. 
Although excited states carry errors propagated from the approximated GS, the scale of these errors is below that of the sampling noise. According to the authors of the original qEOM article \cite{ollitrault_quantum_2020}, for the energy at least, the ratio between the absolute error for the excited state and the GS becomes smaller as the imprecision on the GS increases. In all cases, the excited states are estimated with higher accuracy compared to the GS. 

In our case, we find a similar result, as the excited states measured with infinite sampling are much more precise than the VQE-approximated GS. This can be seen from the red curve in Fig. \ref{f:ethylene}(b), which tends to zero as the temperature increases, where the weight of the approximated GS decreases and so does the error it carries. 
Conversely, in the other cases displayed in the same plot, the trace distance increases for $\beta$ approaching zero, with the main source of error driving the increment being the finite number of measurement sampling. 

The different behaviour in the cases considered demonstrates that the error due to limited sampling is orders of magnitude higher than the error due to the approximated GS. The error in the excitation energy is always below chemical precision \cite{QuantumChemistry2020}, regardless of temperature. While not shown in the plot, our data indicates that the same statement holds true even for a lower number of sampling shots, down to a value of $10^4$ shots.

\begin{figure*}
	\begin{tabularx}{2\columnwidth}{XX}
		(a)
		
		\includegraphics[width=0.45\textwidth]{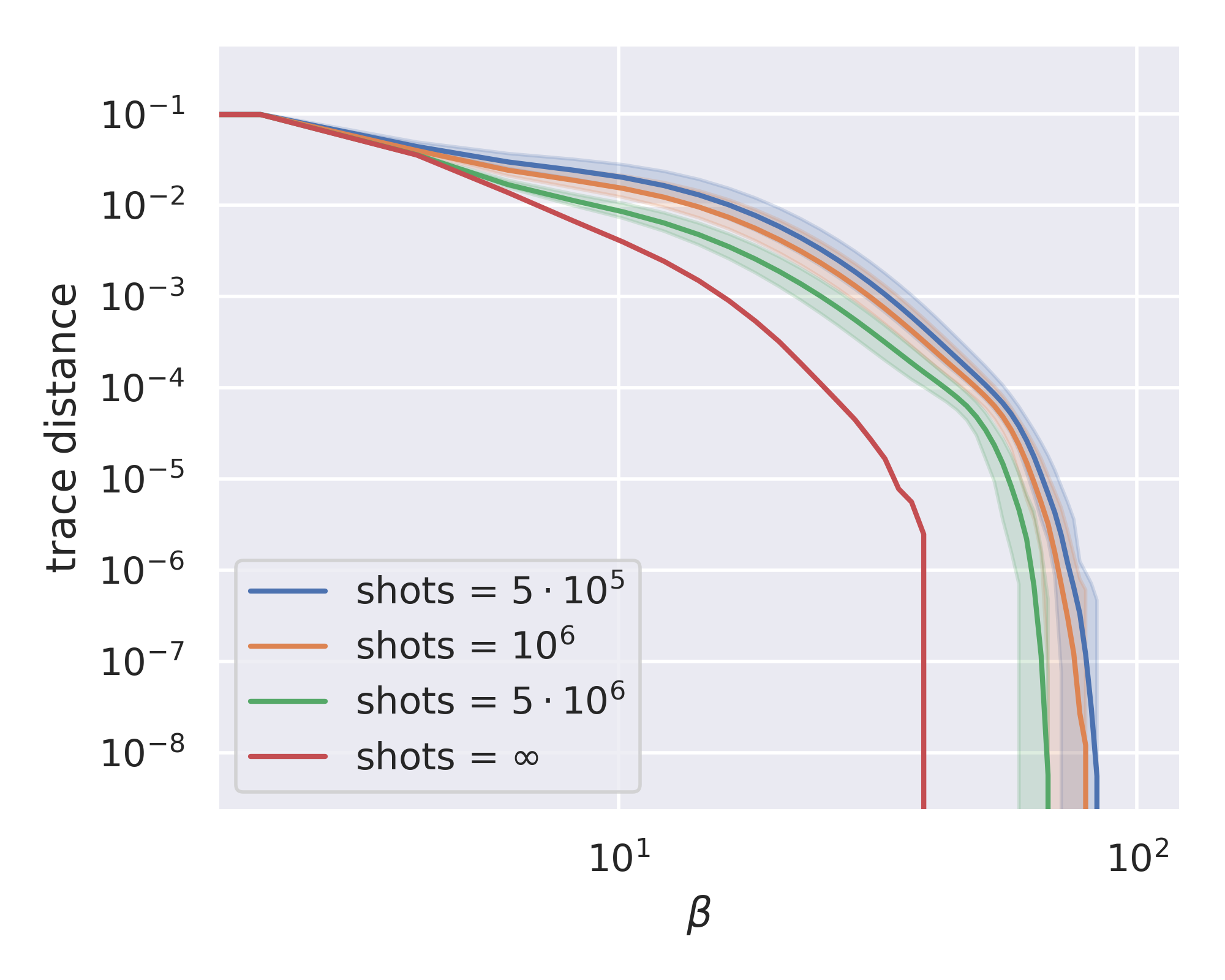}  &
		(b)
		
		\includegraphics[width=0.45\textwidth]{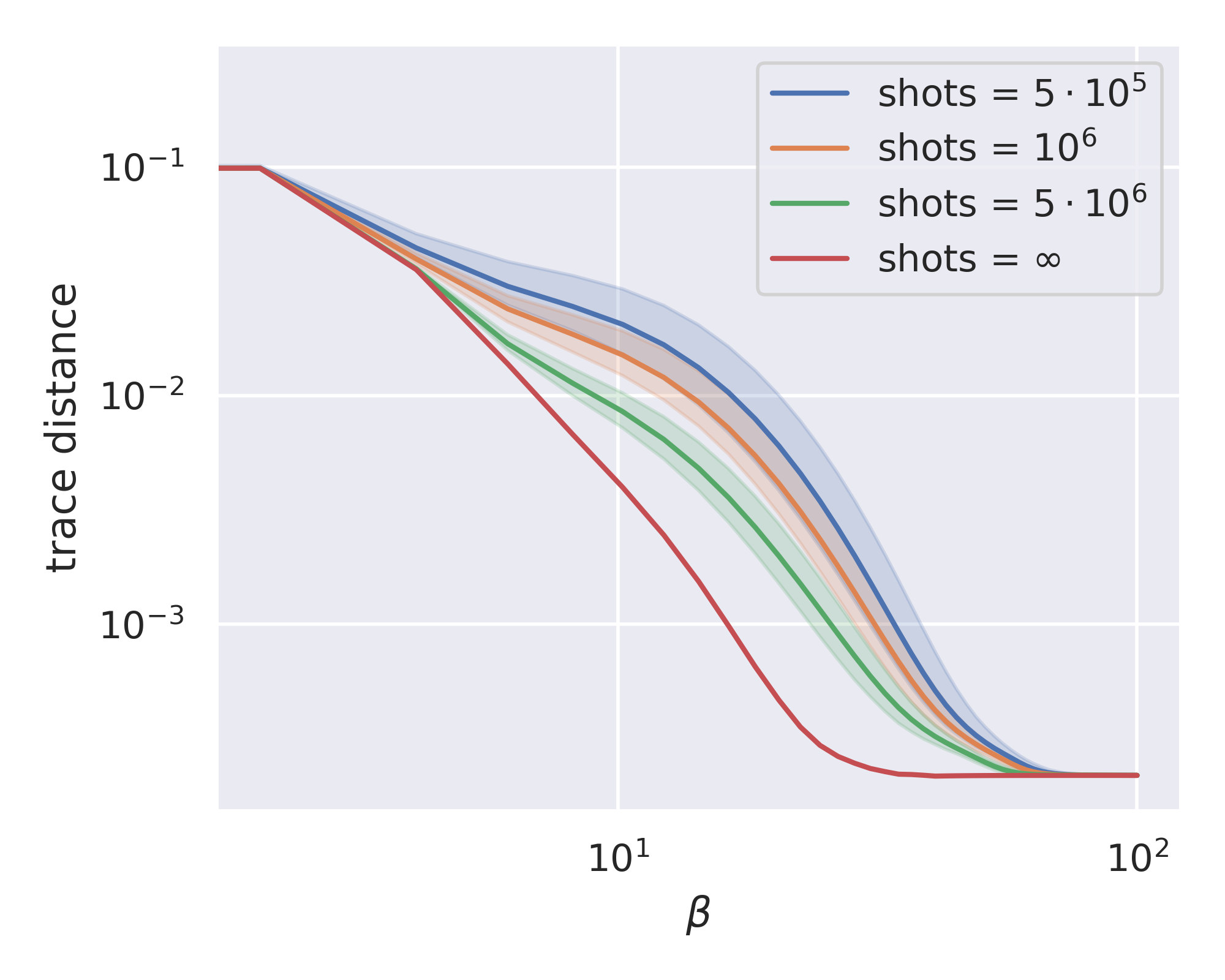}  \\
		(c)
		
		\includegraphics[width=0.45\textwidth]{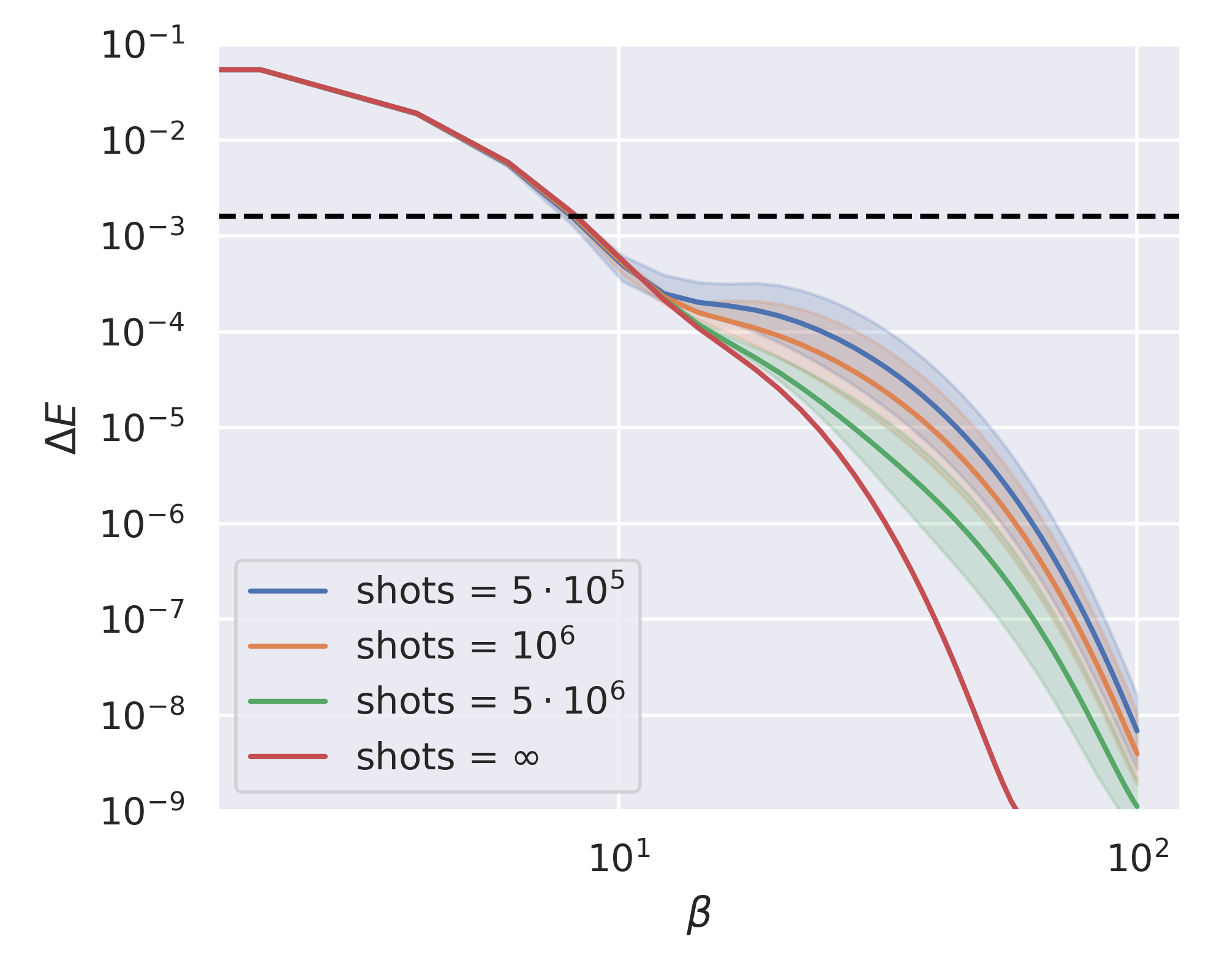}  &
		(d)
		
		\includegraphics[width=0.45\textwidth]{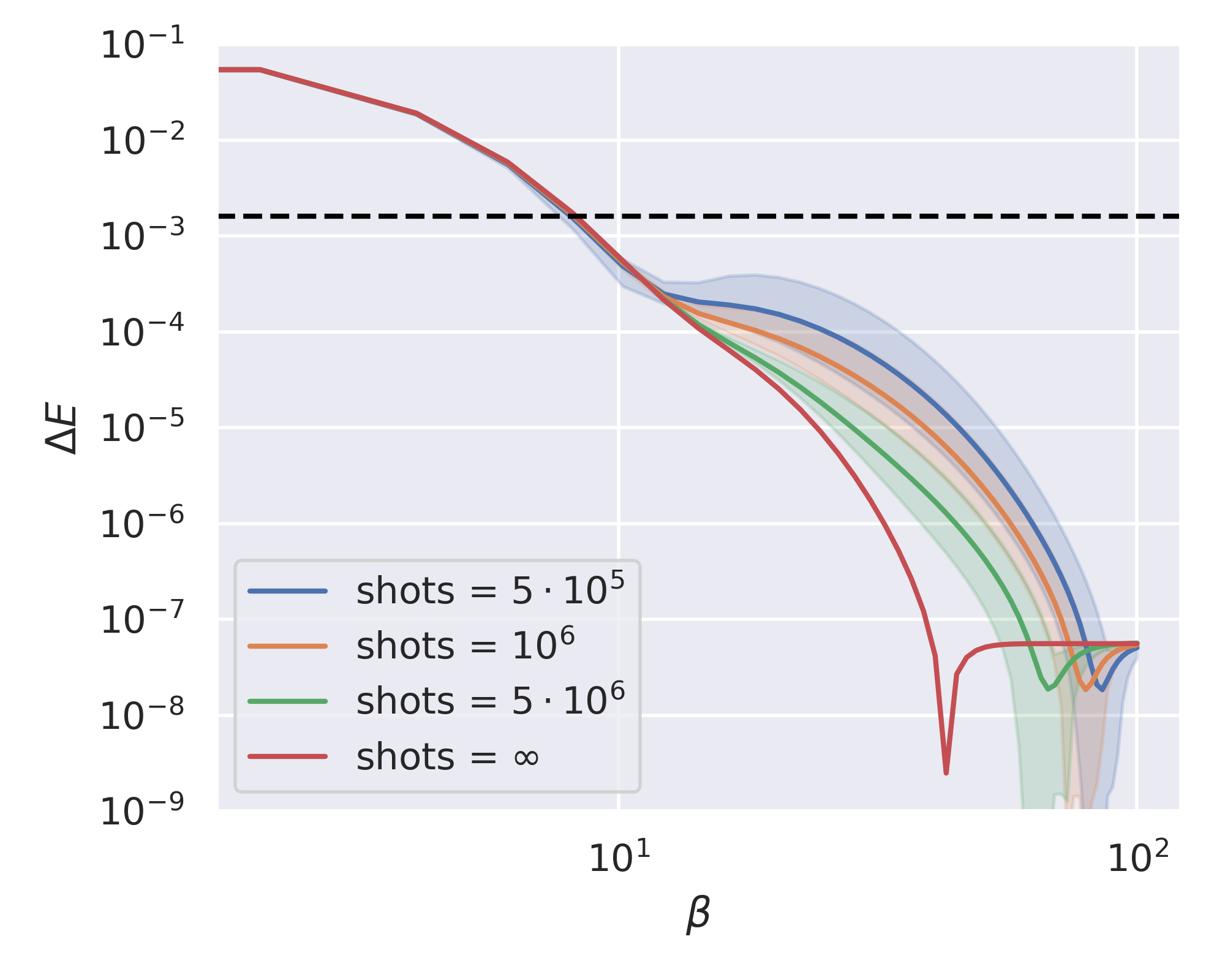}  \\
	\end{tabularx}
	\caption{Butadiene: Top row, trace distance between ideal thermal state and reconstructed thermal state through the qEOM method for different number of shots and starting from the exact ground state (panel a) and from a ground state found through a VQE (panel b). Bottom row, error for the average energy for the same states as above. Dashed black line indicates chemical precision at $1.6\cdot 10^{-3}$ Ha \cite{QuantumChemistry2020}. All simulations run with finite sampling are repeated 100 times, and the spread of results is shown in the figures as an interval with a percentile width of 99.7.} 
	\label{f:butadiene}
\end{figure*}

For butadiene, similar plots are reported in Fig. \ref{f:butadiene}. For the trace distance in the case of an exact GS, see Fig. \ref{f:butadiene} (a), the curve with infinite sampling has a value of zero within numerical accuracy for values of $\beta$ over approximately 40, under which the limited number of excited states limits the accuracy of the approximation. The other curves with limited sampling follow the trend, with evident lower accuracy. In the case of the VQE-approximated GS, the curve with infinite sampling is lower bounded by the accuracy of the GS at higher values of $\beta$.

We report that the instability of the qEOM method, subjected to the type of system it is being applied to, is also influenced by the sampling technique we included in the method.
According to the original paper, when the lowest energy solution is close to the GS, the conditioning of the EOM matrices deteriorates, worsening the approximation of the energies. Our results indicates that this effect is also influenced by other sources of error included in the method. In our case, estimation of the EOM matrices through measurement is also affected by the sampling technique. 

In our study, we report that simulations with a low number of sampling shots, up to $10^4$ shots, were potentially unstable due to the low accuracy of the estimated EOM matrices. 
As for the energy, we included a line for the chemical precision at $1.6 \times 10^{-3}$ Ha \cite{QuantumChemistry2020} on the plot for comparison purposes. Our results seem to indicate that the point where the line for the chemical precision intersects the estimated energy error does not depend significantly on the number of shots or on the choice of the GS. 

These variables may influence the accuracy of the estimated energy, especially in the region below chemical precision. In the low temperature region we see some interplay between shots noise and accuracy, but even with limited sampling the estimated energy always falls within chemical precision. 
At higher values of $\beta$, eigenvalues found using single and double excitation are not enough to accurately estimate the energy, and the approximation falls off. 

For this reason, in the high temperature region, the errors accumulated from the limited sampling and the approximated GS becomes small enough that all curves displayed in Fig.~\ref{f:butadiene} (c) and (d)  overlap. 
While for the energy this may indicate that a higher number of shots may not improve the overall accuracy below a certain temperature, this statement may not hold for all possible quantities of interest, as, unlike the energy, in the trace distance differences in accuracy can be seen at higher values of $\beta$ (in the range of values between $\beta=5$ and $\beta=10$).

\section{Conclusion}
\label{s:conc}
In this work, we investigated the effectiveness of the qEOM algorithm for computing molecular excited states in conjunction with the IC-POVM sampling technique. We applied this method to linear polyenes such as ethylene and butadiene. 
Measuring through IC-POVMs is particularly convenient because it allows for the estimation of several observables at the same time, using the same dataset of raw measurement data. This procedure is much more efficient than the independent estimation of each observable necessary for building the EOM problem, as their number grows rapidly with the system size. It is also clearly more efficient than the computation of the mean values of these observables through classical method, which becomes unfeasible for large molecules.

For small molecules like ethylene, we observe robust numerical agreement between the ideal GS and the approximated GS found through the qEOM at any temperature, and for all the number of shots we have considered, which are meaningful for current near-term computers. The estimated average energy of the thermal state of the molecule remains below chemical precision across all temperatures.

For butadiene, the accuracy of our method depends on the number of measurement shots. For a low number of shots ($10^4$ or lower), we have found numerical instabilities in the solution of the qEOM problem for this molecule. Using $10^5$ shots or more, which is doable on near-term quantum computers, the method becomes stable. In this case we observe a delicate interplay between the number of shots and the accuracy of estimates, particularly at intermediate values of $\beta$. Notably, in this situation, the approximation consistently delivers estimates of energy with precision surpassing chemical precision.

In conclusion, despite the challenges posed by larger molecular systems, our findings underscore the utility of the qEOM algorithm in accurately computing molecular thermal states. As advancements in quantum computing continue, methods that harness the potential of such devices promise to outperform classical computation in a wide class of tasks, including molecular simulations at finite temperature. In this work we showed that the qEOM method can be run more efficiently through the IC-POVM sampling technique, paving the way for further test and research on the matter. Further work may include using the adaptive IC-POVM scheme as described in \cite{GarcíaPérezGuillermo2021LtMA} and testing the effect of noise on the preparation of the GS. Furthermore, simulation for bigger molecules may be run on a quantum hardware.

\begin{acknowledgments} The authors acknowledge helpful discussions with S. Knecht, D. Cavalcanti, G. García-Pérez, and S. Filippov. DM acknowledges financial support from MUR under the “PON Ricerca e Innovazione 2014-2020”.

\end{acknowledgments}

\bibliography{bibliography}

\end{document}